\documentclass[aip,jcp,reprint]{revtex4-2}
\usepackage{amsfonts}
\usepackage{amsmath}
\usepackage{graphicx}
\usepackage[raggedright,nooneline]{subfigure}
\usepackage{tabularx}
\usepackage{dcolumn}
\usepackage{longtable}
\usepackage{tensor}
\usepackage{color}
\usepackage{placeins}
\usepackage{bm}
\usepackage{amssymb}
\usepackage[urlcolor=white]{hyperref}
\providecommand{\U}[1]{\protect\rule{.1in}{.1in}}

\providecommand{\U}[1]{\protect\rule{.1in}{.1in}}
\begin{document}
\title{Refined approach to cellularization: going from Heller's thawed Gaussian approximation to the Herman--Kluk propagator}
\author{Sergey V. Antipov}
\thanks{Current address: Tananaev Institute of Chemistry, Kola Science Centre, Russian Academy of Sciences, Akademgorodok 26a, 184209 Apatity, Russia}
\affiliation{Laboratory of Theoretical Physical Chemistry, Institut des Sciences et Ing\'enierie Chimiques, Ecole Polytechnique F\'ed\'erale de Lausanne (EPFL),
CH-1015 Lausanne, Switzerland}
\author{Fabian Kr\"oninger}
\affiliation{Laboratory of Theoretical Physical Chemistry, Institut des Sciences et Ing\'enierie Chimiques, Ecole Polytechnique F\'ed\'erale de Lausanne (EPFL),
CH-1015 Lausanne, Switzerland}
\author{Ji\v{r}\'{\i} J. L. Van\'i\v{c}ek}
\email{jiri.vanicek@epfl.ch}
\affiliation{Laboratory of Theoretical Physical Chemistry, Institut des Sciences et Ing\'enierie Chimiques, Ecole Polytechnique F\'ed\'erale de Lausanne (EPFL),
CH-1015 Lausanne, Switzerland}
\date{\today}

\begin{abstract}
We present a refined cellularization (Filinov filtering) scheme for the semiclassical Herman--Kluk
propagator, which employs the inverse Weierstrass transform and optimal
scaling of the cell's size with the number of cells, and was previously used
only in the context of the dephasing representation. 
In the new methodology, the sampling density for the cell centers correlates with the cell size, allowing for an effective sampling of the phase space covered by the initial state of the system. 
The main advantage of the presented approach is that, unlike the standard cellularization, it converges to the original Herman--Kluk result in the limit of an infinite number of trajectories and to the thawed Gaussian approximation when a single trajectory is used. 
We illustrate the performance of the refined cellularization scheme by calculating autocorrelation functions and spectra of both integrable and chaotic model systems.

\end{abstract}
\maketitle

\section{Introduction}

\label{sec:intro}

The semiclassical Herman-Kluk
propagator\cite{Heller:1981,Herman_Kluk:1984,Kay:1994,Miller:2001,Kay:2005,Lasser_Sattlegger:2017}
is a well-established approximation to the quantum propagator. 
By associating semiclassical phases to an ensemble of classical trajectories, the Herman--Kluk propagator incorporates quantum effects, such as coherence and zero-point energy, into the dynamics. 
However, the direct application of this approach to large systems is computationally unfeasible due to the oscillatory nature of the integrand involved.

The oscillatory integrand problem has been ameliorated by various methods, such as the prefactor approximations,\cite{Gelabert_Miller:2000,Zhang_Pollak:2004,Issack_Roy:2007,DiLiberto_Ceotto:2016}
time-averaging,\cite{Kaledin_Miller:2003,Buchholz_Ceotto:2016,Buchholz_Ceotto:2018}
hybrid dynamics,\cite{Grossmann:2006,Goletz_Grossmann:2009,Grossmann:2016} and
cellularization.~\cite{Heller:1991, Walton_Manolopoulos:1996} 
The latter approach, which is the focus of the present work, is based on grouping the neighboring trajectories into phase-space cells and using the information along the central trajectory to evaluate contribution from the whole cell approximately analytically.
Mathematically, it is identical to applying the Filinov integral conditioning technique\cite{Filinov:1986} to the Herman--Kluk approximation,\cite{Walton_Manolopoulos:1996} and we use both names interchangeably for the rest of the paper. 
The method and its variant\cite{Makri_Miller:1987, Makri_Miller:1988} have been widely used to improve Monte Carlo statistics\cite{Makri_Miller:1987, Makri_Miller:1988, Walton_Manolopoulos:1996,Herman:1997,Wang_Miller:2001,Spath_Miller:2004,Antipov_Ananth:2015,Church_Ananth:2017,Church_Ananth:2018} 
and to derive new approximate semiclassical methods for molecular dynamics
simulations.\cite{Thoss_Miller:2001,Antipov_Ananth:2015,Church_Ananth:2017}

Despite a number of successful applications, the standard cellularization approach possesses certain problems. 
The width of introduced cells is an arbitrary parameter and for any finite width the calculated result does not converge to the Herman--Kluk propagator even in the limit of an infinite number of trajectories. 
Thus, the convergence with respect to the width of Gaussian cells has to be investigated,\cite{Walton_Manolopoulos:1996} increasing the overall computational cost. 
In this paper, we present a refined cellularization approach which solves both problems. 
The method has a single convergence parameter---the number of trajectories, which determines both the cell width and the sampling density for the cell centers. 
In addition, the refined approach has two appealing properties: (i) it does converge to the Herman--Kluk approximation when the number of trajectories goes to infinity and, (ii) for Gaussian initial states, a single-trajectory result is unique and equal to the thawed Gaussian approximation.\cite{Heller:1975}

The rest of the paper is organized as follows: Section~II gives a brief overview of the standard cellularization approach applied to the Herman--Kluk propagator and presents the refined cellularization (Filinov filtering) method while discussing its behavior in limits of either a single or infinitely many trajectories. 
We numerically demonstrate the key features of the new method by exploring the dynamics of model integrable and chaotic systems in Sec. III, while Sec.~IV provides conclusions.

\section{Theory}

\label{sec:theory}

In the time-dependent approach to spectroscopy,\cite{Heller:1981, book_Tannor:2007, book_Heller:2018} the power spectrum of a
molecule, which provides information on vibrational frequencies, can be
calculated as the Fourier transform
\begin{equation}
P(E)=h^{-1}\int_{-\infty}^{\infty}C(t)e^{iEt/\hbar}dt \label{eq:spec}
\end{equation}
of the wavepacket autocorrelation function (survival amplitude)
\begin{equation}
C(t)=\langle\Psi_{i}|e^{-i\hat{H}t/\hbar}|\Psi_{i}\rangle, \label{eq:corr} 
\end{equation}
where $|\Psi_{i}\rangle$ is a generic reference state and $\hat{H}$ is the nuclear Hamiltonian operator.

\subsection{Herman--Kluk propagator and Filinov filtering}

\label{sec:HK}

Using the semiclassical Herman--Kluk approximation for the time-evolution operator,~\cite{Herman_Kluk:1984,Miller:2001,Kay:2005} the autocorrelation function~(\ref{eq:corr}) can be written as
\begin{equation}
C^{\mathrm{HK}}(t)=h^{-D}\int \ R_{t}(z_{0})e^{iS_{t}(z_{0})/\hbar
}\langle\Psi_{i}|z_{t}\rangle\langle z_{0}|\Psi_{i}\rangle d^{2D}z_{0}. \label{eq:HK}
\end{equation}
Here $D$ is the number of degrees of freedom, $z_{0}:=(q_{0},p_{0})$ and
$z_{t}:=(q_{t},p_{t})$ are $2D$-dimensional vectors of the initial and final
phase-space coordinates of a classical trajectory, $S_{t}(z_{0})$ is the
classical action along that trajectory, and $R_{t}(z_{0})$ is the Herman--Kluk
prefactor
\begin{align}
R_{t}(z_{0})  &  =\det[(\vphantom{\frac{1}{1}}M_{qq}+\gamma^{-1}\cdot
M_{pp}\cdot\gamma\nonumber\\
&  \left.  \left.  -iM_{qp}\cdot\gamma+i\gamma^{-1}\cdot M_{pq}\right)
/2\right]  ^{1/2}, \label{eq:prefactor}
\end{align}
where $M_{\alpha\beta}=\partial\alpha_{t}/\partial\beta_{0}$ are the components of the stability matrix, $|z_{t}\rangle:=|q_{t}p_{t}\rangle$ is the coherent state
centered at $(q_{t},p_{t})$ and described by the wavefunction
\begin{equation}
\langle q|z_{t}\rangle=\left(  \frac{\det\gamma}{\pi^{D}\hbar^{D}}\right)
^{1/4}e^{(-x^{T}\cdot\gamma\cdot x /2+ip_{t}^{T}\cdot x )/\hbar} \label{eq:CS}
\end{equation}
with shifted position $x:=q-q_{t}$;
$\gamma$ is a $D\times D$ real symmetric positive-definite width matrix. 
To simplify the presentation, in the remainder of the paper we assume that the initial state of the system is a coherent state, $|\Psi_{i}\rangle\equiv|z_{i}\rangle$.
The integral over phase space in Eq.~(\ref{eq:HK}) is usually evaluated using Monte Carlo techniques by sampling the initial conditions $(q_{0},p_{0})$ for classical trajectories. 
The most common choice of the sampling function for Eq.~(\ref{eq:HK}) is the Husimi distribution $\rho(z_{0})=|\langle z_{0} |\Psi_{i}\rangle|^{2}$ of the initial state, which permits rewriting the autocorrelation function as the phase-space average
\begin{equation}
C^{\mathrm{HK}}(t)=\left\langle R_{t}(z_{0})e^{iS_{t}(z_{0})/\hbar}
\dfrac{\langle\Psi_{i}|z_{t}\rangle}{\langle\Psi_{i}|z_{0}\rangle
}\right\rangle _{\rho(z_{0})}, \label{eq:HKav}
\end{equation}
where the notation $\left\langle A(z)\right\rangle _{\rho(z)}$ was used for the ratio
\begin{align}
    \left\langle A(z)\right\rangle _{\rho(z)}:=\frac{\int A(z)\rho(z)d^{2D}z}
{\int\rho(z)d^{2D}z}.
\end{align}

Detailed mathematical implementation of the standard Filinov filtering (FF) procedure for the Herman--Kluk propagator has been outlined in Ref.~\onlinecite{Walton_Manolopoulos:1996}. 
Briefly, the Gaussian cells are introduced into Eq.~({\ref{eq:HKav}}) through the normalized Gaussian integral followed by expanding the original integrand about the cell centers, thus accounting for the contribution from each cell approximately analytically. 
The final expression for the correlation function then reads
\begin{equation}
C^{\mathrm{FF}}(t)=\left\langle F_{t}^{\mathrm{FF}}(z_{0},\Sigma)R_{t}
(z_{0})e^{iS_{t}(z_{0})/\hbar}\dfrac{\langle\Psi_{i}|z_{t}\rangle}{\langle
\Psi_{i}|z_{0}\rangle}\right\rangle _{\rho(z_{0})}, \label{eq:FF}
\end{equation}
with the only difference from the original Herman--Kluk approximation (\ref{eq:HKav}) being the filtering factor
\begin{align}
F_{t}^{\mathrm{FF}}(z_{0},\Sigma)  &  =\sqrt{\det\left( 2B^{-1}\cdot\Sigma\right)}e^{A^{T}\cdot B^{-1}\cdot A/(4\hbar)},
\label{eq:filinov}
\end{align}

where the $2D$-dimensional vector $A$ and the $2D\times2D$ matrix $B$ are defined as
\begin{align}
A  &  =M^{T}\cdot\left(  \Sigma_{0}+iJ\right)  \cdot(z_{t}-z_{i}
)\nonumber\label{eq:A}\\
&  +\left(  \Sigma_{0}-iJ\right)  \cdot(z_{0}-z_{i}),\\
B  &  =M^{T}\cdot\Sigma_{0}\cdot M+\Sigma_{0}+2\Sigma. \label{eq:B}
\end{align}

Here
$\Sigma=\hbar\sigma\ \mathrm{Id}_{2D}$ is a matrix whose determinant is inversely proportional to the phase-space volume of the Gaussian cells, $\sigma$ is the filtering parameter, $\Sigma_{0}=\left(
\begin{smallmatrix}
\gamma & 0\\
0 & \gamma^{-1}
\end{smallmatrix}
\right)  $ is the matrix containing the width parameters of the initial coherent state $|z_{i}\rangle$, and $J=\left(
\begin{smallmatrix}
0 & -\mathrm{Id}_{D}\\
\mathrm{Id}_{D} & 0
\end{smallmatrix}
\right)  $ is the standard symplectic matrix. 
Varying the filtering parameter $\sigma$ changes the phase-space volume where dynamics is calculated approximately, which affects the statistical convergence. 
In the limit of $\sigma\rightarrow\infty$, Eq.~(\ref{eq:FF}) reduces to the original Herman--Kluk approximation (\ref{eq:HKav}).

\subsection{Refined cellularization scheme}

\label{sec:RFF}

The refined cellularization scheme was originally introduced for the dephasing representation\cite{Vanicek:2004,Vanicek:2006,Mollica_Vanicek:2011, Wehrle_Vanicek:2011} in Refs.~\onlinecite{Zambrano_Vanicek:2013,Sulc_Vanicek:2012}; here we explain how it can be applied to the Herman-Kluk propagator. 
First, we define a general Gaussian function on phase space as
\begin{equation}\label{eq:cell}
    G_{\Sigma}(z):= \sqrt{\det\left(\Sigma\right)}  e^{-z^{T} \cdot \Sigma \cdot z /(2\hbar)}. 
\end{equation}
Note that $G_{\Sigma}$ satisfies the normalization condition $h^{-D}\int G_{\Sigma}(z) d^{2D}z =1$. 
Contrary to the standard Filinov procedure, we choose the Gaussian cells to be scaled versions of the initial state. 
In particular, the width matrix of each cell is
\begin{equation}
 \Sigma=\lambda^{-2}\Sigma_{0}, \label{eq:cell_width}
\end{equation}
where the scaling parameter $\lambda$ is inversely proportional to the number of cells $N$,
\begin{equation}
\lambda=N^{-\frac{1}{2D}}, \label{eq:scaling}
\end{equation}
and varies in the range $\lambda\in(0,1]$. Such choice of $\lambda$ guarantees that the total phase-space volume occupied by all cells is equal to the volume of the initial state. Next, we use the Gaussian functions~(\ref{eq:cell}) to express the Husimi distribution of the original state through convolution:
\begin{equation}
\rho(z_{0})=h^{-D}\int \ C_{\Sigma}^{\rho}(z)G_{\Sigma}(z_{0}-z) d^{2D}z.
\label{eq:conv}
\end{equation}
The coefficient $C_{\Sigma}^{\rho}(z)$ of this expansion is known as the inverse Weierstrass transform of $\rho(z_{0})$ and for the initial coherent state $|z_{i}\rangle$ it can be evaluated analytically as
\begin{equation}
C_{\Sigma}^{\rho}(z)=G_{\Lambda}(z-z_{i}), \label{eq:cell_weight}
\end{equation}
where 
$\Lambda=(1-\lambda^{2})^{-1}\Sigma_{0}$. 
Inserting the
expansion~(\ref{eq:conv}) into Eq.~(\ref{eq:HK}) and following the standard
Filinov procedure for linearization of the
dynamics\cite{Walton_Manolopoulos:1996} yields the final \emph{refined Filinov
filtered} (RFF) expression
\begin{equation}
C^{\mathrm{RFF}}(t)=\left\langle F_{t}^{\mathrm{RFF}}(z_{0},\lambda
)R_{t}(z_{0})e^{iS_{t}(z_{0})/\hbar}\dfrac{\langle\Psi_{i}|z_{t}\rangle
}{\langle\Psi_{i}|z_{0}\rangle}\right\rangle _{C_{\Sigma}^{\rho}(z_{0})}
\label{eq:RFF}
\end{equation}
for the wavepacket autocorrelation function, where the new filtering factor is
given by
\begin{align}
F_{t}^{\mathrm{RFF}}(z_{0},\lambda)  &  =\sqrt{\det\left(2 Y^{-1}\cdot\Sigma\right)}e^{ X^{T}\cdot  Y
^{-1}\cdot X /(4\hbar)}\label{eq:Rfilinov}
\end{align}
with
\begin{align}
X  &  =M^{T}\cdot\left(  \Sigma_{0}+iJ\right)  \cdot(z_{t}-z_{i}
)\nonumber\label{eq:X}\\
&  -\left(  \Sigma_{0}+iJ\right)  \cdot(z_{0}-z_{i}),
\\& Y=M^{T}\cdot\Sigma_{0}\cdot M-\Sigma_{0} + 2\Sigma. \label{eq:Y}
\end{align}

\begin{figure*}[ptbh]
\includegraphics[width=12cm]{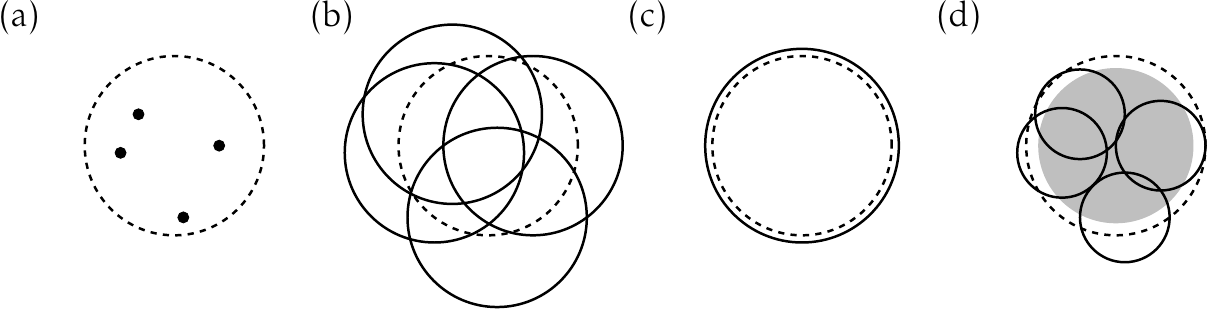}\caption{Comparison of standard and
refined cellularization schemes for $D=1$. Standard sampling: (a) Centers
(dots) of $N=4$ Gaussian functions are sampled from the Husimi function of the
initial state (dashed circle). (b)~Corresponding basis functions (solid
circles) have widths independent of $N$. Proposed sampling: (c) For $N=1$,
there is no freedom in choosing the center and the method agrees with the
thawed Gaussian approximation. (d)~For $N=4$, Gaussian centers are sampled
from the grey disk (of radius $\sqrt{3}/2$ times the radius of the initial
state) and their width is $1/2$ of the initial state's width.}\label{fig:RFF}
\end{figure*}

Despite obvious similarities of Eqs.~(\ref{eq:FF}) and (\ref{eq:RFF}), there
are certain \emph{qualitative} differences between the two cellularization
schemes (Fig.~{\ref{fig:RFF}}). In the original approach, the initial
phase-space coordinates for the centers of the Gaussians are sampled from a
predefined distribution, which is independent of the number $N$ and size
$\Sigma$ of the cells. Therefore, taking the limit $N\rightarrow\infty$ for a
fixed nonzero $\Sigma$ is wasteful since many cells are overlapping
[Fig.~\ref{fig:RFF}(b)]. Moreover, in this limit the standard method does not
converge to the Herman--Kluk result and the error depends on the accuracy of
the approximate treatment of contributions from trajectories within each cell,
i.e., on the cell size. Thus, the convergence of the calculation has to be
explored with respect to both the number and size of cells.

The refined cellularization scheme alleviates all these problems---as the
number of cells increases, their size decreases, while the sampling density
for the cell centers becomes broader [Fig.~\ref{fig:RFF}(d)]. In the limit
$N\rightarrow\infty$, the cells reduce to points and the sampling density,
$C_{\Sigma}^{\rho}(z)$, converges to the Husimi distribution $\rho(z)$ of the
initial wavefunction $\Psi_{i}$. In this limit, $F_{t}^{\mathrm{RFF}
}\rightarrow1$ and the refined Filinov filtering approach~(\ref{eq:RFF})
converges to the original Herman--Kluk correlation function~(\ref{eq:HKav}),
which is the crucial advantage of the method.

For the single cell, $N=1$, there is no freedom in choosing the position of
its center, since the sampling density~(\ref{eq:cell_weight}) reduces to a
delta-function,
\begin{equation}
C_{\Sigma}^{\rho}(z)\rightarrow h^{D}\delta(z-z_{i}),
\end{equation}
and the center coincides with that of the initial state, while the cell
occupies the same volume as the initial Husimi distribution. Thus, in contrast
to the standard cellularization approach, the refined Filinov filtering
provides a natural choice of parameters for the single cell. 
Furthermore, it was shown by Grossmann~\cite{Grossmann:2006} that performing the
cellularization expansion of the Herman--Kluk integrand around the center of
the initial wavepacket amounts to making the thawed Gaussian approximation of
Heller.\cite{Heller:1975,Heller:2006,Wehrle_Vanicek:2014,Begusic_Vanicek:2018}
Thus, the proposed refined cellularization scheme applied to the Herman--Kluk wavefunction reduces to another
well-known semiclassical approximation in the limit of a single cell.
In Appendix~\ref{Appendix:Sec_2}, we additionally show the equality of the refined Filinov filtered autocorrelation function $(C^{\rm RFF})$ in the limit of $N=1$ trajectory and the thawed Gaussian wavepacket $(C^{\rm TGWP})$ autocorrelation function.

\section{Numerical examples}

\label{sec:results}

To illustrate the numerical performance of the new cellularization scheme, we
calculate the autocorrelation functions and Franck--Condon spectra in both
integrable and chaotic model systems.

For an integrable system, we consider a two-dimensional model of the collinear
NCO molecule including the ground ($X^{2}\Pi$) and first excited ($A^{2}
\Sigma^{+}$) electronic states.~\cite{Li_Buenker:1993, Zambrano_Vanicek:2013}
This example is chosen as a typical molecular system---anharmonic but not
fully chaotic; a detailed description of the model can be found in Ref.~\onlinecite{Zambrano_Vanicek:2013}.

\begin{figure}
\centering
\includegraphics[scale = 0.9]{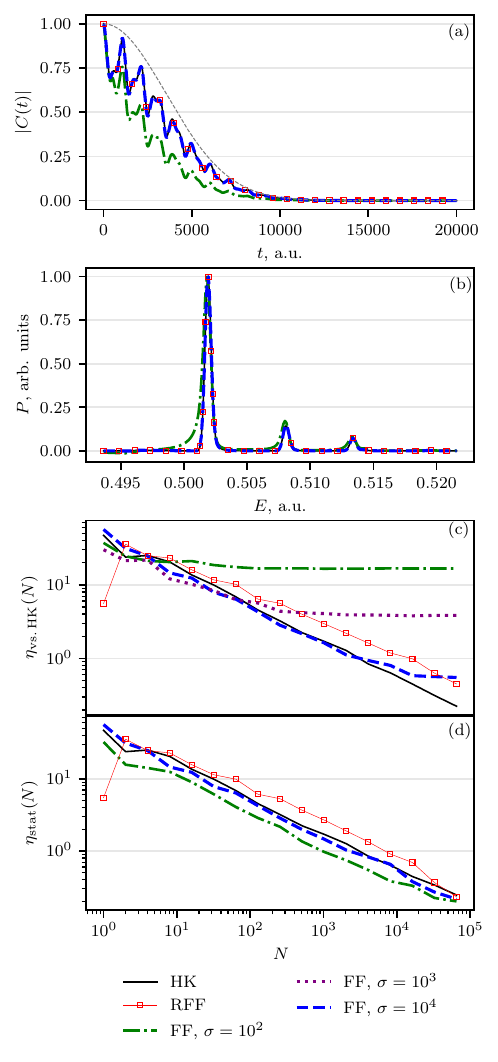}
\caption{The collinear NCO model: Comparison of the refined
Filinov filtering (RFF), Herman--Kluk (HK) and standard Filinov filtering (FF)
results  calculated for different values of the filtering parameter $\sigma$.
(a) Time correlation functions calculated with $N=2^{17}\approx1.3 \times 10^{5}$
trajectories multiplied by the damping function (indicated by a dashed line).
(b) Corresponding Franck--Condon spectra. All spectra were scaled so that the highest
intensity is $1$. (c) Convergence error $\eta_{\rm vs.\,HK}$ of the damped correlation
functions to the fully converged Herman--Kluk result 
(obtained with $N_{\mathrm{ref}}=2^{24}\approx1.68\times10^{7}$ trajectories) as a function of the number
of trajectories $N$. 
Each line was produced by averaging the error over $10$ independent simulations. For clarity, the standard Filinov result for $\sigma=10^3$ is only shown in panel (c).
(d) Statistical convergence $\eta_{\rm stat}$ of the damped correlation functions to $C_{N_{\rm ref}}$ ($N_{\rm ref}=2^{17}$) within each method. 
Each line was produced by averaging the error over $10$ independent simulations. 
}\label{fig:NCO}
\end{figure}

The initial coherent state is chosen to match the ground vibrational state of
the $X^{2}\Pi$ electronic state [i.~e. a Gaussian centered at $q_i=(3.585,2.246)$ and  $p_i=(0,0)$ with width matrix $\gamma =\left(
\begin{smallmatrix}
116.0704 & -30.3084\\
30.3084 & 118.5828
\end{smallmatrix}
\right) $ and mass $m=\mathrm{diag}(17021.13,12505.97)$] and is evolved on the $A^{2}\Sigma^{+}$
surface with step size $4$ a.\,u. up to time $T=56000$.  Figure~\ref{fig:NCO}a compares the calculated time correlation
functions revealing a very good agreement between the refined Filinov
filtering and Herman--Kluk results; the original cellularization approach
shows substantial deviations if the Gaussian cells are not small enough (i.e.,
filtering parameter $\sigma$ is not sufficiently large). The latter difference
translates to the spectra: the corresponding peaks are significantly broadened
and have a different lineshape compared to the Herman--Kluk result
[Fig.~\ref{fig:NCO}(b)]. Again, the refined cellularization and Herman--Kluk
approaches are in remarkable agreement with each other.

We also investigate the convergence of the three methods to the converged Herman--Kluk result, which is
quantified by the $L^{2}$ error
\begin{equation}
\eta_{\mathrm{vs.\,HK}}(N):=||C_{N}-C^{\rm HK}_{N_{\mathrm{ref}}}|| \label{eq:L2error}
\end{equation}
achieved for $N<N_{\mathrm{ref}}$ trajectories, where $||f||^{2}:=\int
_{0}^{T}|f(t)|^{2}dt$, $C_{N}$ is the autocorrelation
function evaluated with a given method and $N$ trajectories, and $C^{\rm HK}_{N_{\rm ref}}$ is the converged Herman--Kluk result (obtained with $N_{\rm ref}$ trajectories)

Figure~\ref{fig:NCO}(c) confirms that
the refined Filinov filtering approach continues to gradually improve with
increasing number of trajectories, while in the standard procedure a
convergence plateau is reached for a given value of the filtering parameter
$\sigma$.

Moreover, inspection of the single trajectory $(N=1)$ results in Fig.~\ref{fig:NCO}(c) shows that the new
cellularization scheme agrees with the fully converged Herman--Kluk result much more than do the $1-$trajectory results obtained with the Herman--Kluk and standard Filinov methods. 
This comes at no surprise since the
presented approach prescribes a unique, judicious choice for the center and
size of the single cell, whereas in the other two methods the center is sampled randomly, leading, in general,
to a poor description of the initial wavepacket.

In Fig.~\ref{fig:NCO}(d), we show the purely statistical error 
\begin{align}\label{EQ:statistical_convergence}
    \eta_{\rm stat}(N) = \vert\vert C_N - C_{N_{\rm ref}} \vert\vert
\end{align}
of each method as a function of $N$. 
The statistical error of the refined cellularization 
is initially much lower than the error of the original Herman--Kluk approach and increases rapidly when going from $N=1$ to $N=2$ trajectories (because the initial conditions are not uniquely defined when $N\neq 1$). 
Then, the statistical error of the refined cellularization remains parallel to the original Herman--Kluk error until it converges to it from above as the number of trajectories increases.
As expected, a smaller filtering parameter for the standard filtering approach results in a lower statistical error. 
For a fixed filtering parameter $\sigma$, the statistical convergence error remains parallel to and usually lower than the Herman--Kluk error.
Because $C^{\rm FF}$ converges to $C^{\rm HK}$ in the limit of $\sigma\rightarrow \infty$,  the statistical convergence error $\eta_{\rm stat}(N)$ of the Filinov filtered approach converges to the one from the Herman--Kluk approach by continuity.

To illustrate the utility of the refined Filinov filtering procedure for
chaotic systems, we consider a two-dimensional quartic oscillator with potential
\begin{equation}
 V(q) = \frac{1}{20} \left( q_{1}^{4} + q_{2}^{4} \right) + \frac{1}{2} q_{1}^{2} q_{2}^{2}.
\label{eq:quartic}
\end{equation}
The initial coherent state is centered at 
$q_{i} = (0,4)$ and $p_{i} = (4,0)$
with width parameters 
$\gamma = \mathrm{diag}(1,1)$ and masses $m = \mathrm{diag}(1,1)$.
The parameters of the system and initial coherent state are
chosen such that the dynamics is highly chaotic. Therefore, we used a small step size of $0.01$ a.~u.

\begin{figure}
\centering
\includegraphics{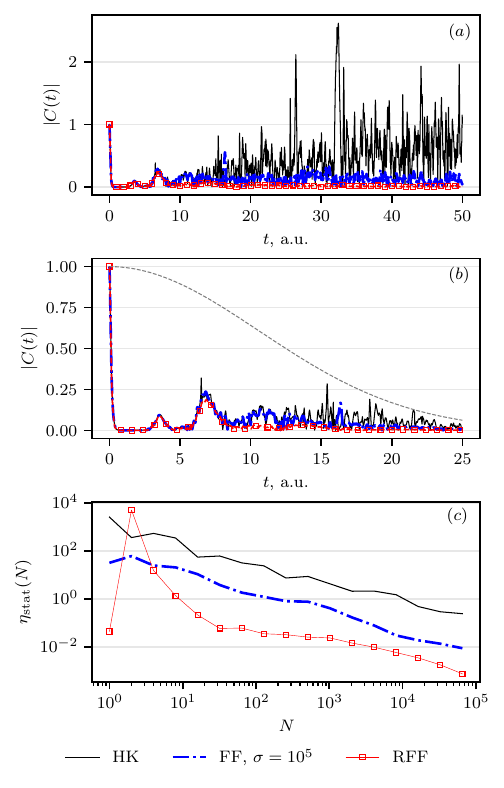}
\caption{The chaotic quartic oscillator: Comparison of the refined Filinov filtering (RFF), standard Filinov filtering (FF) and Herman--Kluk (HK) results. 
(a) Undamped time correlation function calculated with $N= 2^{17}\approx1.3\times10^{5}$ trajectories. 
(b) Short-time behavior of correlation functions from panel (a)
multiplied by a damping function (shown by a dashed grey curve). 
(c) Convergence error $\eta_{\rm stat}$ of the damped correlation function to a reference solution $C_{N_{\rm ref}}$ ($N_{\rm ref}=2^{17}$) within each method. Each line was produced by averaging the error over $10$ independent simulations.}\label{fig:quartic}
\end{figure}

Figure~\ref{fig:quartic}(a) compares the time-correlation functions calculated
using the Herman--Kluk, standard Filinov, and refined Filinov methods. Due to the
chaotic nature of the dynamics, fully converging the Herman--Kluk calculation
is virtually impossible, since a few chaotic trajectories can result in a very
large prefactor, which explains the unphysically large $[\vert C(t) \vert >1]$ oscillations observed for
$t>20$. The problem is partially fixed by the cellularization approaches---the
filtering factors in Eqs.~(\ref{eq:FF}) and (\ref{eq:RFF}) damp the unphysical
oscillatory structure of the correlation function at long times.

Since chaotic systems provide a stress-test for the Herman--Kluk approximation
and its variants, it is instructive to investigate the purely statistical convergence
within each method [using Eq.~(\ref{EQ:statistical_convergence})]. 
For this, we consider correlation functions multiplied by
an arbitrary damping function [Fig.~\ref{fig:quartic}(b)]; this approach is
often used in theoretical spectroscopy to reproduce the effect of the line
broadening in the experiment. 
While all considered methods show a poor
convergence rate [Fig.~\ref{fig:quartic}(c)], the refined cellularization approach
shows considerably smaller absolute errors. Increasing the number of cells
decreases their size and, therefore, the filtering strength. As a consequence,
using just a small number of trajectories allows to effectively damp the
chaotic oscillatory structure at long times and provides a good description of
the short-time correlation function. Thus, the method can be effectively used
to simulate the short-time dynamics, before the chaos develops fully.

\section{Conclusions}

We have introduced the refined cellularization of the Herman--Kluk wavepacket autocorrelation function. In this method, the number of trajectories determines the width of the cells and the density from which the centers of the cells are sampled. 
While the standard Filinov filtering procedure introduces a normalized Gaussian integral with a fixed width, the refined cellularization introduces cells through convolution of the Husimi distribution with a cell whose size is determined by the number of cells.
This choice has the advantage that in the limit of infinitly many cells, the method converges to the original Herman--Kluk correlation function, and in the limit of a single cell, the cell's center is deterministic and the resulting correlation function  reduces to the autocorrelation function obtained with the thawed Gaussian wavepacket approximation.

We tested this refined filtering method and compared it to the original, unfiltered Herman--Kluk approach as well as to the standard Filinov filtering.
In the fairly anharmonic NCO model, the statistical convergence of the refined cellularization shows a modest improvement over the Herman--Kluk approach.
In the chaotic quartic oscillator, the refined cellularization results in a remarkable improvement over the Herman--Kluk approach and standard Filinov filtering, accelarating the convergence by a few orders of magnitude.

Finally, the refined cellularization approach has approximately the same computational cost per trajectory as the Herman--Kluk and standard Filinov filtering methods. The total cost is primarily determined by the propagation cost of a single classical trajectory, including its classical action and stability matrix. 

\label{sec:summary}

\acknowledgments The authors acknowledge the financial support from the
European Research Council (ERC) under the European Union's Horizon 2020
research and innovation programme (Grant Agreement No. 683069-MOLEQULE).




\appendix

\section{Refined cellularization}
Here we derive the expression (\ref{eq:RFF}) from the main text for the refined cellularized wavepacket autocorrelation function.
Using the convolution \eqref{eq:conv} and changing the order of integration, the Herman--Kluk autocorrelation function (\ref{eq:HK}) can be re-expressed as
\begin{align}
    C^{\rm HK}(t) = \langle h(y_{0}) \rangle_{C_{\Sigma}^{\rho}(y_{0})}
\end{align}
with the integrand
\begin{align}
    h(y_{0}) = h^{-D} \int R_{t}(z_{0}) e^{iS_{t}(z_{0})} \frac{\langle \psi_{i} \vert z_{t} \rangle}{ \langle \psi_{i} \vert z_{0} \rangle} G_{\Sigma}(\delta z) \,dz_{0}
\end{align}
and $\delta z = z_{0} - y_{0}$.

Applying the Filinov filtering procedure, i.~e., Taylor expanding $S_{t}$, $z_{t}$, and $R_{t}$ to the second, first, and zeroth order around $y_0$, respectively, we obtain
\begin{align}
    h(y_{0})  &\approx h^{-D}  \left(\det{\Sigma}\right)^{1/2} R_{t}(y_{0}) 
    \\&\int \exp{\left[\frac{i}{2\hbar}g_{t}(z_{0},y_{0})-\frac{f_{t}(z_{0},y_{0})}{4\hbar}\right]} \,d^{2D}z_{0} 
\end{align}
with the real-valued functions 
\begin{align}
\begin{split}
f_{t}(z_{0},y_{0}) &=  2\delta z^{T} \cdot \Sigma \cdot \delta z
\\&-(y_{0} +\delta z -z_{i})^{T} \cdot \Sigma_{0} \cdot (y_{0}+\delta z -z_{i})  
\\ &+ \left(z_{i}-y_{t}-M\cdot \delta z \right)^{T} \cdot \Sigma_{0} 
\\& \cdot (z_{i}-y_{t}-M\cdot \delta z), 
\end{split}
\\
\begin{split}
g_{t}(z_{0},y_{0}) &= \left(p_{i}+p_{t} + M_{pq}\cdot \delta q + M_{pp} \cdot \delta p  \right)^{T}
\\& \cdot \left(q_{i}-q_{t} - M_{qq}\cdot \delta q - M_{qp} \cdot \delta p\right)
\\ &  + (p_{0} +\delta p + p_{i})^{T} \cdot (q_{0}+\delta q -q_{i})
\\ &  + 2S_{t}(y_{0}) + 2 \nabla S_{t}(y_{0}) ^{T} \cdot \delta z  
\\& + \delta z^{T} \cdot \begin{pmatrix} M_{pq}^{T} \cdot M_{qq} & M_{pq}^{T} \cdot M_{qp} \\ M_{qp}^{T} \cdot M_{pq} & M_{pp}^{T} \cdot M_{qp} \end{pmatrix} \cdot \delta z.
\end{split}
\end{align}

Using the symplecticity relation $M^T \cdot J \cdot M = J$, 
we obtain the simplified form
\begin{align}
\begin{split}
h(y_{0}) & =  h^{-D} \left(\det{\Sigma}\right)^{1/2} 
R_{t}(y_{0}) e^{iS_{t}(y_{0})/\hbar} \frac{\langle \psi_{i} \vert y_{t} \rangle}{\langle \psi_{i} \vert y_{0} \rangle}
\\ &\times \int e^{\left[-(\delta z)^{T} \cdot Y \cdot \delta z - 2 X^{T} \cdot \delta z \right] / 4\hbar} \, d^{2D}z_{0}
\end{split}
\end{align}
with $Y= M^{T} \cdot \Sigma_{0} \cdot M -\Sigma_{0} + 2\Sigma$ and $X=M^{T} \cdot (\Sigma_{0} + iJ)(y_{t}-z_{i}) - (\Sigma_{0}+iJ)(y_{0}-z_{i}).$
Integration over $z_{0}$ yields the final expression (\ref{eq:RFF}) for $C^{\rm RFF}(t)$ from Sec.~\ref{sec:RFF}.

\section{Connection of the refined cellularization to the thawed Gaussian approximation}\label{Appendix:Sec_2}

When $N=1$, we show that $C^{\rm RFF}(t)$ is equal to the overlap 
$\langle \psi_{0} \vert \psi_{t} \rangle$
of two thawed Gaussian wavepackets
\begin{align}
\begin{split}
    &\langle q \vert \psi_{t} \rangle = \frac{1}{(\pi\hbar)^{D/4}\sqrt{\det Q_{t}}}
    \\& \times \exp\left[ \frac{i}{\hbar} \left( \frac{1}{2} x^T \cdot C_{t} \cdot x + p_{t}^T \cdot x + S_{t} \right) \right],
\end{split}
\end{align}
where $C_{t} = P_{t} \cdot Q_{t}^{-1}$ is the product of the ``Hagedorn'' matrices\cite{Lasser_Lubich:2020, Vanicek:2023} $Q_{t} = M_{qq}\cdot Q_{0} + M_{qp} \cdot P_{0}$, $ P_{t} = M_{pq} \cdot Q_{0} + M_{pp} \cdot P_{0}$ with initial conditions $Q_{0} = \gamma^{-1/2}$, $P_{0} = i\gamma^{1/2}$, $(q_{t=0}, p_{t=0}) = (q_{i}, p_{i})$, and $M = \partial z_{t} / \partial z_{i}$.

If $N=1$, $\lambda = 1$, $Y= M^T \cdot \Sigma_{0} \cdot M + \Sigma_{0}$
and  $C^{\rho}_{\Sigma}(z_{0})$ reduces to the delta-function $h^D \delta (z-z_{i})$.
Therefore, $X=M^T \cdot (\Sigma_{0}+ iJ) \cdot (z_{t} -z_{i})$, where $z_{t}$ is the initial condition $z_{i}$ propagated up to time $t$ and $M=\partial z_{t} / \partial z_{i}$.
The wavepacket autocorrelation function simplifies to
\begin{align} \label{EQ:C_Rff_appendix}
    C^{\rm RFF}(t) = F_{t}^{\rm RFF}(z_{i},1) R_{t}(z_{i}) e^{iS_{t}(z_{i})/\hbar} \langle \Psi_{i} \vert z_{t} \rangle.
\end{align}
Next, we decompose $Y$ into the matrix product
\begin{align}
    &Y
      = 
    -iM^T \cdot \begin{pmatrix}
        -iP_{0} & 0 \\  iQ_{0} &Q_{0}
    \end{pmatrix} \cdot \begin{pmatrix}
        \mathrm{Id}_D &  i\mathrm{Id}_D/2 \\ 0 & \mathrm{Id}_D
    \end{pmatrix}
    \\&
    \cdot \begin{pmatrix}
        P_{0}\cdot Q_{t} + Q_{0} \cdot P_{t}, & 0
        \\ 0, &  P_{0} \cdot \overline{Q}_{t} - Q_{0} \cdot \overline{P}_{t}
    \end{pmatrix}
    \\& \cdot
    \begin{pmatrix}
        \mathrm{Id}_D, & -i\mathrm{Id}_D/2 \\ 0, & \mathrm{Id}_D
    \end{pmatrix}\cdot\begin{pmatrix}
       -i P_{0}, & 0 \\ -P_{0}, & Q_{0} 
    \end{pmatrix},
\end{align}
where $\mathrm{Id}_D$ denotes the $D$-dimensional identity matrix.
Hence, one can deduce the equality
\begin{align}\label{EQ:Appendix_B7}
    \sqrt{\det(2Y^{-1} \cdot \Sigma)} R_{t}(z_{i}) 
    =\frac{(2i)^{D/2}}{\sqrt{\det(P_{0}\cdot Q_{t} + Q_{0} \cdot P_{t})} }
\end{align}
from the symplecticity of $M$ and from the equation
\begin{align}
   [R_{t}(z_{i})]^2 = (2i)^{-D} \det(P_{0} \cdot \overline{Q}_{t} - Q_{0} \cdot \overline{P}_{t}). 
\end{align}
The symmetric matrix $X^T \cdot Y^{-1} \cdot X$ in the exponent of the refined filtering factor $F_{t}^{\rm RFF}(z_{i},1)$ is equal to
\begin{align}
    &(\Sigma_{0}+iJ)^T \cdot M \cdot Y^{-1} \cdot M^T \cdot (\Sigma_{0} + iJ)
    \\& = 
    i \begin{pmatrix}
        -i C_{0}, & i\mathrm{Id}_D \\ -i\mathrm{Id}_D, & iC_{0}^{-1}
    \end{pmatrix}
    \cdot 
    \begin{pmatrix}
        Q_{t}, & i\overline{Q}_{t}/2
        \\ P_{t}, & i\overline{P}_{t}/2
    \end{pmatrix}
    \\&
    \cdot \begin{pmatrix}
        Q_{0} \cdot (C_{t}+C_{0}) \cdot Q_{t}, & 0
        \\ 0, & Q_{0} \cdot (C_{0} - \overline{C}_{t}) \cdot \overline{Q_{t}}
    \end{pmatrix}^{-1}
    \\& \cdot \begin{pmatrix}
        -i P_{0}, & -i Q_{0}
        \\ 0, & 0
    \end{pmatrix}
    \\&
    = 
    i\begin{pmatrix} 
        C_{-}\cdot C_{+}^{-1} \cdot C_{0}, 
        & C_{-}\cdot C_{+}^{-1}
        \\  C_{+}^{-1}\cdot C_{-},
        & C_{0}^{-1}\cdot C_{-} \cdot C_{+}^{-1}
    \end{pmatrix}  \label{EQ:Appendix_B11}
\end{align}
with $C_{+}= C_{t} + C_{0}$ and $C_{-} = C_{t} - C_{0}$.
Combining Eqs.~(\ref{EQ:C_Rff_appendix}), (\ref{EQ:Appendix_B7}) and (\ref{EQ:Appendix_B11}), we obtain
\begin{align}\label{EQ:Overlap_gaussians_appendix}
\begin{split}
    &C^{\rm RFF}(t) = \frac{(2i)^{D/2}}{\sqrt{\det(P_{0}\cdot Q_{t} + Q_{0} \cdot P_{t})} }  e^{iS_{t}(z_{i})/\hbar}
    \\& \times e^{ \frac{i}{2\hbar} \left[ (z_{t}-z_{i})^T \cdot \Xi \cdot (z_{t}-z_{i}) + (p_{i} + p_{t})^T \cdot (q_{i} - q_{t}) \right] },
\end{split}
\end{align}
where the width matrix
\begin{align}
    \Xi &= \frac{i}{2}\Sigma_{0} - \frac{i}{2} X^T \cdot Y^{-1} \cdot X
    \\& =
    \begin{pmatrix}
        2 [C_{0} - C_{0} \cdot C_{+}^{-1} \cdot C_{0} ] ,
        & 
        C_{-}\cdot C_{+}^{-1}
        \\
        C_{+}^{-1} \cdot C_{-},
        &
        -2 C_{+}^{-1}
    \end{pmatrix}   
\end{align}
was obtained using the Sherman-Morrison-Woodbury formula.\cite{Higham:2002}

The right-hand side of Eq.~(\ref{EQ:Overlap_gaussians_appendix}) is equal to the overlap $\langle \psi_{0}  \vert \psi_{t} \rangle$ of two Gaussians with width matrices $C_{0}$, $C_{t}$ and phase-space centers $z_{0}$, $z_{t}$, respectively.\cite{Bergold_Lasser:2022}
Because the equations of motion of the trajectory $z_{t}$, its classical action $S_{t}$ and stability matrix $M$ are equivalent to those under a local harmonic approximation of the potential, and because the initial conditions are $z_{t=0}=z_{i}$ and $C_{0} = i\gamma$, Eq.~(B2) is equivalent to the overlap $\langle \psi_{i} \vert z_{t} \rangle $ of the Gaussian initial state with the thawed Gaussian wavepacket.

\bibliographystyle{aipnum4-2}
\bibliography{CellularHK}

\end{document}